\documentclass[PRL,aps,twocolumn,showpacs,preprintnumbers,amsmath,amssymb,superscriptaddress,nofootinbib]{revtex4-1}
\usepackage[dvips]{graphicx}
\usepackage{epsf}
\usepackage{amsmath}
\usepackage{amssymb}
\usepackage{amsfonts}
\usepackage{times}
\usepackage[usenames]{color}
\usepackage[normalem]{ulem}
\usepackage[T1]{fontenc}
\usepackage[dvipsnames]{xcolor}

\usepackage{graphicx}% Include figure files
\usepackage{dcolumn}% Align table columns on decimal point
\usepackage{bm}% bold math
\usepackage{color}

\begin{document}

\title{Effective Dark Matter Halo catalog in $f(R)$ gravity}
%\title{Dark Matter Halo properties in modified gravity}

\author{Jian-hua He}
\email[Email address: ]{jianhua.he@brera.inaf.it}
\affiliation{INAF-Observatorio Astronomico, di Brera, Via Emilio Bianchi, 46, I-23807, Merate (LC), Italy}

\author{Adam~J.~Hawken}
\affiliation{INAF-Observatorio Astronomico, di Brera, Via Emilio Bianchi, 46, I-23807, Merate (LC), Italy}

\author{Baojiu Li}
\affiliation{Institute for Computational Cosmology, Department of Physics, Durham University, Durham DH1 3LE, UK}

\author{Luigi Guzzo}
\affiliation{INAF-Observatorio Astronomico, di Brera, Via Emilio Bianchi, 46, I-23807, Merate (LC), Italy}

\begin{abstract}
We introduce the idea of {\it effective} dark matter halo catalog in $f(R)$ gravity, which is built using the {\it effective} density field. Using a suite of high resolution N-body simulations, we find that the dynamical properties of halos, such as the distribution of density, velocity dispersion, specific angular momentum and spin, in the effective catalog of $f(R)$ gravity closely mimic those in the $\Lambda$CDM model. Thus, when using effective halos, an $f(R)$ model can be viewed as a $\Lambda$CDM model.  This effective catalog therefore provides a convenient way for studying the baryonic physics, the galaxy halo occupation distribution and even semi-analytical galaxy formation in $f(R)$ cosmologies.
\end{abstract}

\maketitle
%{\bf Introduction.}~ It is well established that the Universe is currently undergoing a period of accelerated expansion \cite{1,WMAP,planck,BAOm}. The leading interpretation for this accelerated expansion is the cosmological constant in the context of general relativity. However, this standard paradigm suffers from the coincidence problem as well as the cosmological constant problem (see, e.g., Ref.~\cite{sean} for a review). An alternative interpretation for the cosmic acceleration is modified theory of gravity. General relativity might not be accurate on cosmological scales. The gravity might follow different rules on such a large scale. Chameleon $f(R)$ theory (see. Ref.~\cite{frreview,review_Tsujikawa} for reviews) is one of such theories that admit a reasonable background expansion history of the Universe~\cite{frmodel,HuS} as well as an efficient screening mechanism in the local perturbed space-time~\cite{Khoury,frsim}. Testing such theories of gravity can help us understand the nature of the cosmic acceleration. It is therefore one of the main tasks of upcoming cosmological surveys such as Euclid mission~\cite{Euclid}.

{\bf Introduction.}~ It has become well established that the Universe is currently undergoing a period of accelerated expansion \cite{1,2,3,4,5,6,WMAP,planck,BAOm}. The predominant explanation for this phenomenon is that it is driven by a non-zero cosmological constant, $\Lambda$, in the framework of General Relativity (GR). Together with the assumption that most of the matter in the Universe is cold and dark (non baryonic), this forms the current $\Lambda$CDM standard cosmological model. There are however theoretical arguments as to why such an explanation should be disfavoured, such as the discrepancy between the value of the cosmological constant measured astronomically and that predicted by quantum field theory (see, e.g., Ref.~\cite{sean} for a review). An alternative explanation %to the accelerated expansion
is that GR might not be accurate on cosmological scales and so some modification to it may be necessary to match observations.

A popular family of modified gravity models come under the umbrella of chameleon $f(R)$ gravity (so called because it replaces the Ricci scalar $R$ in the Einstein-Hilbert action in GR with some function $f(R)$; see, e.g., Ref.~\cite{frreview,review_Tsujikawa} for recent reviews). This function introduces an effective cosmological constant, which allows the Universe to expand in a way to match observations \cite{frmodel,HuS}, as well as an extra scalar degree of freedom, which mediates a fifth force. However, it also contains an efficient screening mechanism which can suppress this fifth force %is not felt
in high density environments \cite{Khoury,frsim}, therefore mimicking GR in environments such as our solar system and the early universe.

Placing constraints on such modified theories of gravity can help us understand a lot about the nature of the cosmic acceleration. This is one of the main tasks of upcoming cosmological surveys such as the Euclid mission~\cite{Euclid}. In order to make competitive forecasts for constraining $f(R)$ with upcoming surveys we need mock galaxy catalogs which resemble the observations we expect the satellite to make. These are needed to investigate systematic errors which could impact observations. To do this, we must not only eventually be able produce large N-body simulations covering many ${\rm Gpc}^3h^{-3}$ but also to populate these dark matter simulations with galaxies.

In order to produce mock galaxy catalogs in modified cosmologies, it is necessary to have an understanding of galaxy formation and evolution. Although galaxies are extremely complicated objects and many details of the physical processes still remain poorly understood even within the $\Lambda$CDM paradigm, encouraging progress has been made in recent years. State-of-the-art hydrodynamical simulations such as {\sc Illustris}~\cite{illustris} and {\sc Eagle}~\cite{EAGLE}, with proper modelling of subgrid astrophysics, are able to reproduce galaxy properties that are in good agreement with observations. Although hydrodynamical simulations can faithfully follow the ``gastrophysics'' in a gravitational field during the hierarchical process of structure formation, such simulations are computationally expensive and high resolution simulations in $f(R)$ gravity are not available currently. An alternative approach is to use semi-analytical galaxy formation models \cite{semi}, which derive galaxy properties from dark matter simulations. Processes such as the cooling of gas, the formation of stars, feedback effects, and galaxy mergers closely relate to the properties of their host halos (e.g., halos mass, velocity dispersion and merger history). Although semi-analytical models usually contain free parameters, their predictions are found to be in reasonable agreement with observations and the models are well motivated by underlining physics. It is therefore {of particular interest} to study semi-analytical galaxy formation models in $f(R)$ gravity.

However, modifications to gravity increase the complexity of galaxy formation. In $f(R)$ gravity the properties of halos depend not only on their mass but also on their level of screening. For instance, the velocity dispersion is radically different in unscreened halos in $f(R)$ gravity compared to halos of equivalent mass in $\Lambda$CDM. Consequently, the virial temperature of gas in these halos is higher than in $\Lambda$CDM  ~\cite{fr_volker}. It therefore follows that a halo which has assembled under enhanced gravitational forces %which are enhanced
may have altered astrophysics and it should not be assumed that the processes which govern galaxy formation are the same in such a halo.
%For instance, in $f(R)$ cosmologies, the velocity dispersion of an unscreened halo has a significant enhancement compared to a halo with the same mass in the $\Lambda$CDM model due to the effective enhancement of gravitational forces. As a consequence, the virial temperature of gas in these halos in $f(R)$ gravity will be higher than that in the $\Lambda$CDM model~\cite{fr_volker}. Therefore we expect the processes governing the formation of galaxies in unscreened halos in $f(R)$ gravity to be different as well.
%The processes of the formation of galaxy in unscreened halos in $f(R)$ gravity therefore are expected to be different from those in the $\Lambda$CDM model as well.

In order to overcome these difficulties, in this paper we introduce the idea of the {\it effective halo catalog}, which is built using the {\it effective} density field in $f(R)$ gravity. We shall show that the dynamical properties of halos in this catalog closely resemble those in $\Lambda$CDM dark matter halos.

{\bf Setup.} The formation of large-scale structure in $f(R)$ gravity is governed by the modified Poisson equation,
\begin{equation}
\nabla^2\phi=\frac{16\pi G}{3}\delta \rho-\frac{1}{6}\delta R\quad,\label{poissonfr}
\end{equation}
as well as an equation for the scalar field $f_{R}$,
\begin{equation}
\nabla^2\delta f_R=\frac{1}{3c^2}[\delta R - 8\pi G\delta \rho]\quad,\label{frpoisson}
\end{equation}
where $\phi$ is the gravitational potential, $\delta f_R\equiv f_R(R)-f_R(\bar{R})$, $\delta R\equiv R-\bar{R}$, and $\delta \rho\equiv\rho-\bar{\rho}$. The overbar denotes the background values of quantities, and $\nabla$ is the derivative with respect to physical coordinates. Combining Eq.~(\ref{poissonfr}) and Eq.~(\ref{frpoisson}), it follows that
\begin{equation}
\nabla^2\phi_L=4\pi G\delta \rho \label{poissonN}\quad,
\end{equation}
where
\begin{equation}
\phi_L\equiv\phi+\frac{c^2\delta f_R}{2}\quad,\nonumber\label{N_phi}
\end{equation}
is the lensing potential. The gravitational potential $\phi$ is felt by massive particles and is therefore the potential associated with the dynamical properties of halos and the processes of galaxy formation in $f(R)$ gravity.

We ran a suite of high-resolution N-body simulations using the {\sc ecosmog} code~\cite{ECOSMOG},  itself based on the publicly available N-body code {\sc ramses}~\cite{RAMSES}, to solve Eqs.~(\ref{poissonfr}, \ref{frpoisson}). We studied an $f(R)$ model which exactly reproduces the $\Lambda$CDM background expansion history~\cite{frmodel}. Our simulations have a box size of $L_{\rm box}=64 h^{-1}{\rm Mpc}$ and contain $N = 256^3$ particles. The background cosmology matches the Planck \cite{planck} best-fit $\Lambda$CDM model ($\Omega_b^0=0.049, \Omega_c^0=0.267, \Omega_d^0=0.684, h=0.671, n_s=0.962$, and $\sigma_8=0.834$). Initial conditions, at a redshift of $z=49$, were generated using the {\sc mpgrafic} package~\cite{inicon}. Fourteen simulations were run in total, one realisation for $f(R)$ models with $f_{R0}=-10^{-6}$ and $f_{R0}=-10^{-5}$, and five for $f_{R0}=-10^{-5.5}$ (where $f_{R0}$ is the present value of ${\rm d}f/{\rm d}R$). For each $f(R)$ simulation we ran a $\Lambda$CDM one with the same initial conditions as a control.

{\bf Effective halo catalog.}~We define an effective density field $\delta \rho_{\rm eff}$ so that the modified Poisson equation, Eq.~(\ref{poissonfr}), in $f(R)$ gravity can be cast into the same form as Eq.~(\ref{poissonN})
\begin{equation}
\nabla^2\phi=4\pi G \delta \rho_{\rm eff}\quad,\label{eff_poisson}
\end{equation}
where $\delta \rho_{\rm eff}\equiv(\frac{4}{3}-\frac{\delta R}{24\pi G\delta \rho})\delta \rho\,.$
We identify halos in simulations using the {\it true} density field $\delta \rho$ and the {\it effective} density field $\delta \rho_{\rm eff}$, respectively.
The halo radius $R_{\rm h}$ is defined as the radius of a sphere within which the average density, $\bar{\rho}_h$, %inside the sphere
is $\Delta_{\rm h}$ times the mean density, $\bar{\rho}_m$. The total mass inside the halos is %given by
$$M_{\rm h}=\frac{4\pi}{3} R_{\rm h}^3\Delta_{\rm h} \bar{\rho}_m\quad.$$
We modified the {\sc Amiga} Halo Finder ({\sc Ahf}) \cite{AHF} to identify dark matter halos and remove unbound particles taking into account the modification of gravity.
Throughout this work we take $\Delta_{\rm h}=328$ and limit our study to halos containing more than $400$ particles.
We call the catalog of halos identified using the {\it true} density field $\delta \rho$ the {\it standard} catalog. In contrast, we call the catalog of halos identified using the {\it effective} density field $\delta \rho_{\rm eff}$ the {\it effective} catalog. The {\it standard} and {\it effective} catalogs are two {\it different} catalogs. In each catalog, a halo has a well defined lensing mass $M_{L}\equiv\int \delta \rho(\vec{x}){\rm d}V$ and dynamical mass $M_{D}\equiv\int \delta \rho_{\rm eff}(\vec{x}){\rm d}V$. However, there is not a one-to-one correspondence between the masses in the two halo catalogs and there are three aspects of differences between the two catalogs: given the same halo radius, the lensing mass in the {\it standard} catalog is slightly different from that in the {\it effective} catalog; the positions of centers of halos in the {\it effective} catalog are different from those in the {\it standard} catalog; the number counts of halos in the two halo catalogs are totally different as well. Further, it should be noted that $M_{L}$ and $M_{D}$ as defined above are not dependent on the shape of the halo. The dynamical mass $M_{D}$ can be calculated accurately by our definition without the approximation that halos are spherical (e.g., in Refs.\cite{zlk2011, Fabian}).

{\bf Scaling relations.}~After defining the {\it standard} and {\it effective} catalogs, we investigate the relationship between the mass and velocity dispersion, $\sigma_v^2$, of halos in these two catalogs. The virial temperature of gas in virilized gaseous halos is related to the velocity dispersion via a power law, which also applies to $f(R)$ gravity \cite{fr_volker}. We can thus infer the virial temperature of gas in dark matter halos by studying the $M$-$\sigma_v^2$ relationship.
The 3D velocity dispersion of a dark matter halo in the halo-rest frame is defined by $$\sigma_v^2\equiv\frac{1}{N_p}\sum_{i}(\vec{v}_i-\vec{v}_h)^2\quad,$$
where $\vec{v}_h$ and $\vec{v}_i$  are the halo and particle velocities, respectively, and $N_p$ is the number of particles inside the halo.

\begin{figure*}
\includegraphics[width=5in,height=4.5in]{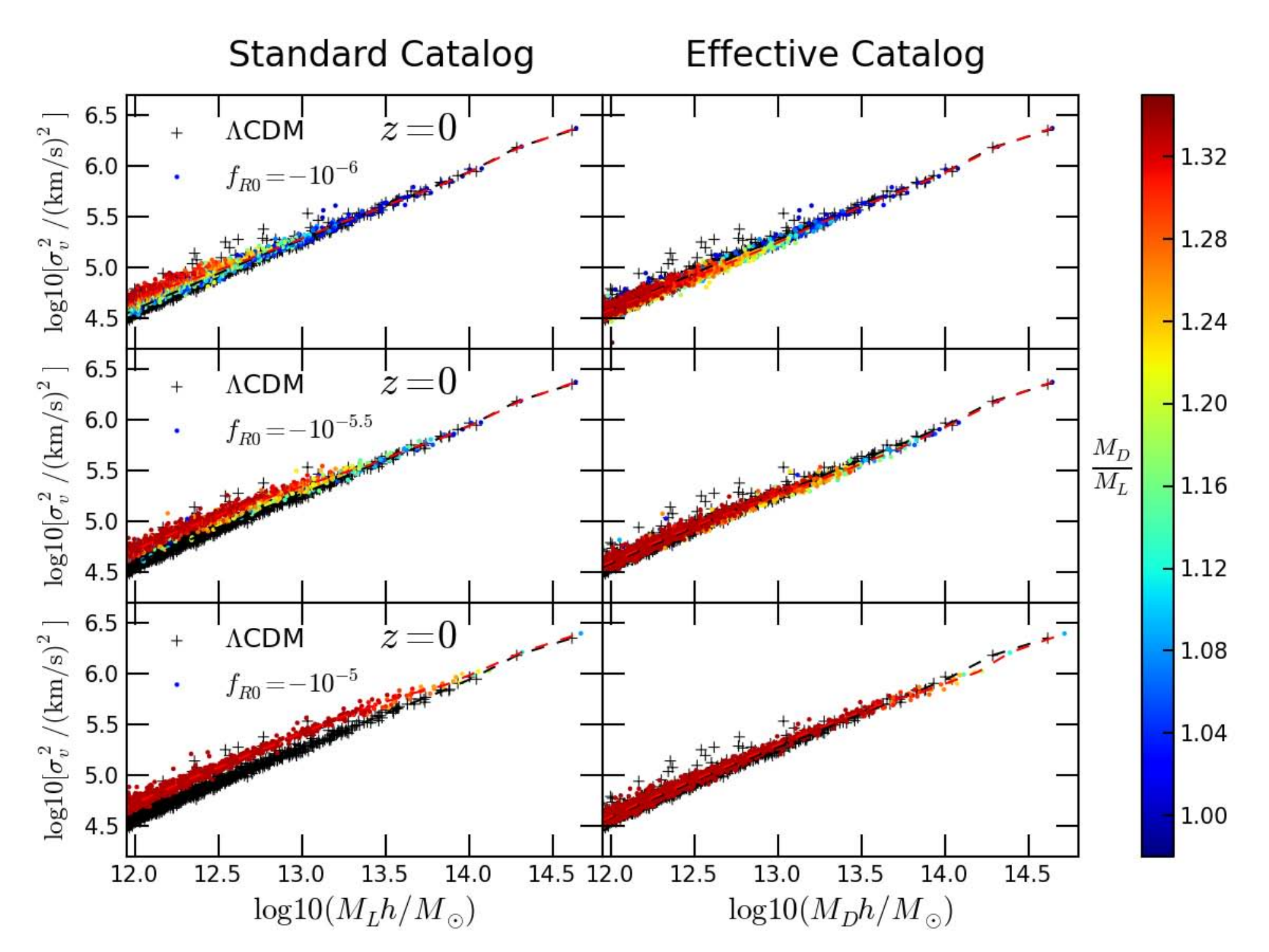}
\caption{The scaling relation of velocity dispersions with respect to halo mass. The points represent $f(R)$ models. The color represents the ratio $M_{D}/M_{L}$. The black crosses represent the $\Lambda$CDM model. The red and black dashed lines are the averaged values. In the left panels, velocity dispersions in $f(R)$ gravity do not scale as a power-law with the lensing mass. The scaling also depends on the screening. In the right panels, the scalings in $f(R)$ gravity are the same as that in the $\Lambda$CDM model.
}\label{figone}
\end{figure*}

Figure \ref{figone} shows the $M$-$\sigma_v^2$ relation for $f(R)$ models with $f_{R0}=-10^{-6},-10^{-5.5},-10^{-5}$ at $z=0$.
In the left-hand panels, the mass used is the lensing mass $M_L$ from the {\it standard} catalog; in the right-hand panels, the mass is the dynamical mass $M_D$ from the {\it effective} catalog. The points represent $f(R)$ halos and the color indicates their level of screening, with the ratio $M_{D}/M_{L}$ illustrated in the color bar to the right. The black crosses represent the halos in the $\Lambda$CDM simulations, and the red and black dashed lines represent the mean values.
In the {\it standard} catalog (left-hand panels), we can see that the velocity dispersions of the well-screened halos (blue) overlap with $\Lambda$CDM halos of equivalent mass. For unscreened halos (red), which are in general less massive, the %relationship between mass and velocity dispersion
$M$-$\sigma_v^2$ relationship is different, with a $\sqrt{4/3}$ enhancement in the velocity dispersion compared to the $\Lambda$CDM case.
However, when we plot the velocity dispersion against the dynamical mass in the {\it effective} catalog (right-hand panels), %we find that the relationship between $M_D$ and $\sigma_v^2$
the $M$-$\sigma_v^2$ relationship is the same as in $\Lambda$CDM for all halos.
%In the left panels, in terms of the lensing mass, the velocity dispersions of the well screened halos (blue) overlap with the results from the $\Lambda$CDM simulation. The velocity dispersions for the unscreend halos (red) have $\sqrt{4/3}\approx1.15$ enhancement than that in the $\Lambda$CDM case~\cite{Fabian,fr_volker}. The $M$-$\sigma_v^2$ scaling relation in terms of the lensing mass obviously depends on the screening levels. However, if we turn to the dynamic mass in the {\it effective} catalog, as is shown in the right panels in Fig.~\ref{figone}, we could find that the dynamic mass $M_D$ has the same scaling relation with $\sigma_v^2$ as that in the $\Lambda$CDM model.
\begin{figure}
\includegraphics[width=3.8in,height=3.8in]{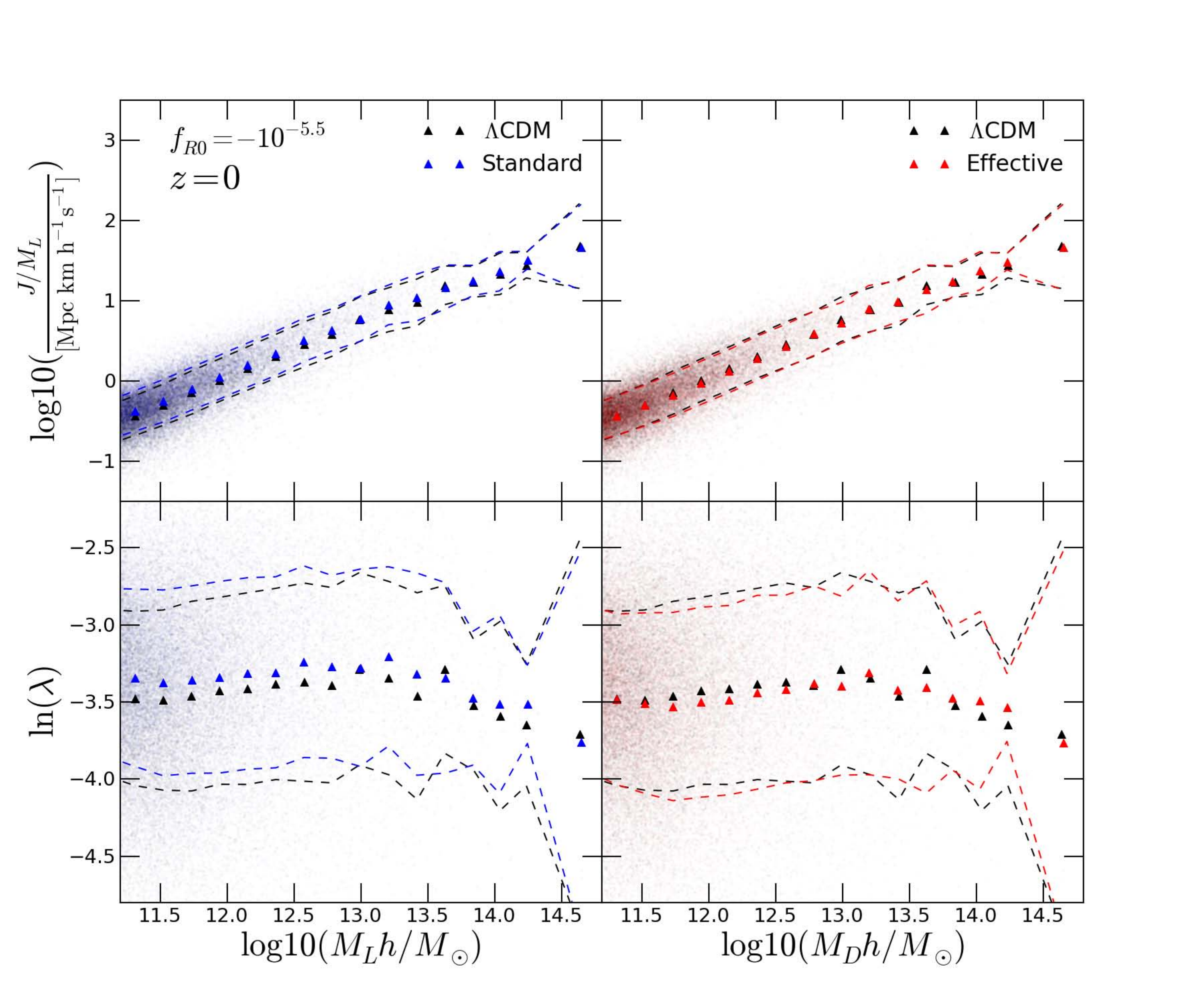}
\caption{Upper panels: the scaling relation of halo mass with the specific angular momentum $j=J/M_L$. The black triangles represent the mean values in each mass bin for the $\Lambda$CDM halos. The blue and red triangles represent the mean values in each mass bin for $f(R)$ halos in the {\it standard} and {\it effective} catalogs, respectively. The dashed lines indicate the $1\sigma$ scatter. Lower panels: the logarithm of the spin parameter $\ln(\lambda)$ as a function of the halos mass.
}\label{figtwo}
\end{figure}

Next, we investigate the relationship between halo mass and angular momentum,
which is defined by
\begin{equation}
\vec{J} \equiv \sum_i^{N_p} m_i\Delta\vec{r}_i\times \Delta\vec{v}_i\quad,
\end{equation}
where $\Delta\vec{r}_i$ and $\Delta\vec{v}_i$ are the position and velocity of the $i \hbox{-} {\rm th}$ particle relative to the mean value of the halo. $m_i$ is the {\it true} ({\it inertial}) mass of the $i \hbox{-} {\rm th}$ particle. The magnitude of the angular momentum is defined as $J\equiv(J_x^2+J_y^2+J_z^2)^{1/2}$. In the literature, %the angular momentum is usually presented in terms of
people often use the specific angular momentum, $j$ $\equiv$ $J/M$, where $M$ is the {\it inertial} mass of %dark matter
halos. Therefore we use the lensing mass $M_L$ in the definition of $j$ in $f(R)$ gravity.

The upper panels of Fig.~\ref{figtwo} show the scaling relation of the specific angular momentum $j$ relative to the mass of halos for the $f(R)$ model with $f_{R0}=-10^{-5.5}$ at $z=0$. The triangles represent the mean values in each mass bin. In the {\it standard} catalog, $f(R)$ gravity boosts the mean value of $j$ by a factor of $\sqrt{4/3}$ for completely unscreened halos ($M<10^{12} M_{\odot}$)~\cite{frspin}, though this is far outweighed by the large scatters around the mean. {In the {\it effective} catalog, $f(R)$ gravity slightly lowers the mean value of $j$ with a small relative difference %of the average value
$\Delta \bar{j}/\bar{j}_{\Lambda{\rm CDM}}\approx3\%$ within the mass bin $[10^{11.5},10^{13.5}]M_{\odot}$.

%Compared to the impact on halo velocity dispersions, the fifth force of $f(R)$ gravity apparently has less of an impact on the angular momentum of halos due to the large scatter of $j$}.

In the literature, the spin parameter is usually defined as\cite{spin},
\begin{equation}
\lambda \equiv \frac{j}{\sqrt{2}VR}\quad,\label{defspin}
\end{equation}
where $j$ is the specific angular momentum and
$V$ is the circular velocity at radius $R=R_{\rm h}$.
The lower panels of Figure~\ref{figtwo} show %the logarithm of the spin parameter
$\ln(\lambda)$ as a function of the halo mass. In $\Lambda$CDM, $\lambda$ depends weakly on the halo mass~\cite{cole}. Within the mass bin $[10^{11.5},10^{13.5}]M_{\odot}$, the mean value of %the spin parameter
$\lambda$ is very small, $\bar{\lambda}\approx 0.0336$ (compared to, say, a self-gravitating and rotationally-supported disk in which $\lambda\approx 0.4$) %but the scatter is
with a relatively large scatter, $\sigma_{\ln \lambda}\approx 0.617$. The %small spin parameter
smallness of $\lambda$ indicates that dark matter halos are mainly supported by random motions of their particles rather than by coherent rotation.

In $f(R)$ gravity, $\lambda$ also weakly depends on the halo mass. In the {\it standard} catalog, we use the definition of $\lambda$ as
\begin{equation}
\lambda \equiv \frac{J}{M_L\sqrt{2GM_LR}}\quad,\label{spin}
\end{equation}
to be consistent with the one used in the literature(e.g. Ref.~\cite{frspin}). The averaged %value of the
spin parameter is significantly boosted by the fifth force as $\bar{\lambda}\approx 0.0367$, with a relative difference $\Delta \lambda/\bar{\lambda}_{\Lambda{\rm CDM}}\approx 9.2\%$ within the mass bin $[10^{11.5},10^{13.5}]M_{\odot}$.

In the {\it effective} catalog, the circular velocity $V$ at radius $R$ is defined using the dynamical mass as $V^2 \equiv GM_D/R\,$ (see Eq.~(\ref{eff_poisson})). The spin parameter $\lambda$, therefore,
\begin{equation}
\lambda \equiv \frac{J}{M_L\sqrt{2GM_DR}}\quad.\label{spin}
\end{equation}
The average %value of the
spin parameter is very close to %that in
the $\Lambda$CDM %model
value, $\bar{\lambda}\approx 0.0326$, with a small relative difference of $\Delta \lambda/\bar{\lambda}_{\Lambda{\rm CDM}}\approx 3\%$. Both {\it standard} and {\it effective} catalogs have approximately the same size of scatter, at $\sigma_{\ln \lambda}\approx 0.617$.

\begin{figure}
\includegraphics[width=3.8in,height=3.8in]{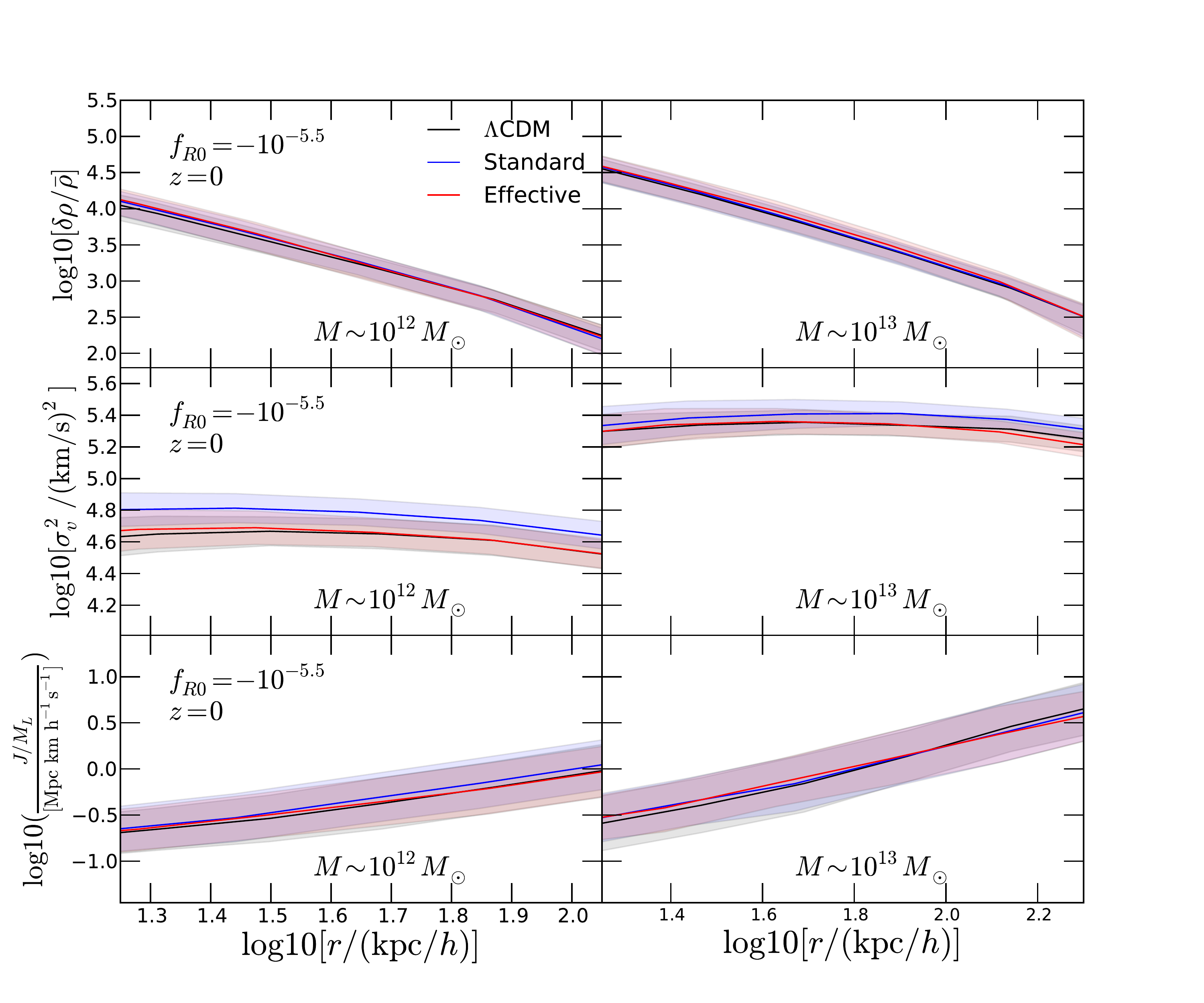}
\caption{Top panels: the density profiles for different mass bins. Middle panels: the velocity dispersion profiles. Bottom panels: $j=J/M_L(r<R)$ within a sphere as the function of radius. The solid lines are the averaged values. The shaded regions represent the $1\sigma$ scatter.}\label{figfour}
\end{figure}

{\bf Profiles.} We now turn %our investigation
to the halo profiles of density, velocity dispersion, and specific angular momentum. %profiles of halos.
Again, we focus our study to the $f(R)$ model with $f_{R0}=-10^{-5.5}$ at $z=0$. We choose two different mass bins in which most dark matter halos are unscreened, $0.95-1.05\times \{10^{12},10^{13}\}M_{\odot}$ (see Fig.~\ref{figone}).
We only consider profiles at $r>10$ ${\rm kpc}/h$ since the accuracy of the halo profiles below this radius is affected by the limited resolution of our simulations.
%Due to the limited resolution of simulations, we only consider the profiles at $r>10 {\rm kpc}/h$.

The velocity dispersion as a function of the radius is defined as $$\sigma_v^2(r)\equiv\frac{1}{\Delta N_p}\sum_{i\in\Delta r}(\vec{v}_i-\vec{v}_h)^2\, ,$$
in which $\Delta N_p$ is the number of particles within the spherical shell $\Delta r$ at a given radius $r$.

The top panels of Fig.~\ref{figfour} show the density profiles for halos in the two %different
mass bins. The solid lines represent the mean values and the shaded regions represent the $1\sigma$ scatter. We can see that the density profiles of halos in $f(R)$ gravity and the $\Lambda$CDM model are indistinguishable in $1\sigma$ range of scatter. On the other hand, when we look at the velocity dispersion profiles (middle panels), we can see that
in the {\it standard} catalog there is
an enhancement in $f(R)$ gravity compared to $\Lambda$CDM. However, we can see that the velocity dispersion profiles are almost identical in both the {\it effective} and $\Lambda$CDM catalogs.
Finally, in the bottom panels of Fig.~\ref{figfour}, we show the specific angular momentum $j(<r)$ %within a sphere
as a function of distance from the halo center. The halos from the {\it effective} catalog and the $\Lambda$CDM model show a good agreement in the distribution of $j$.

{\bf Halo mass function.}~In Fig. \ref{massfunc}, we show the mass function of the effective and standard halos compared to that of the $\Lambda$CDM model. It is clear that the effective halos show more significant enhancement in their abundance than that of the standard halos. This result indicates that the statistics of standard halos may seriously underestimate the impact of $f(R)$ gravity on galaxy clustering.
\begin{figure}
\includegraphics[width=3.6in,height=3.2in]{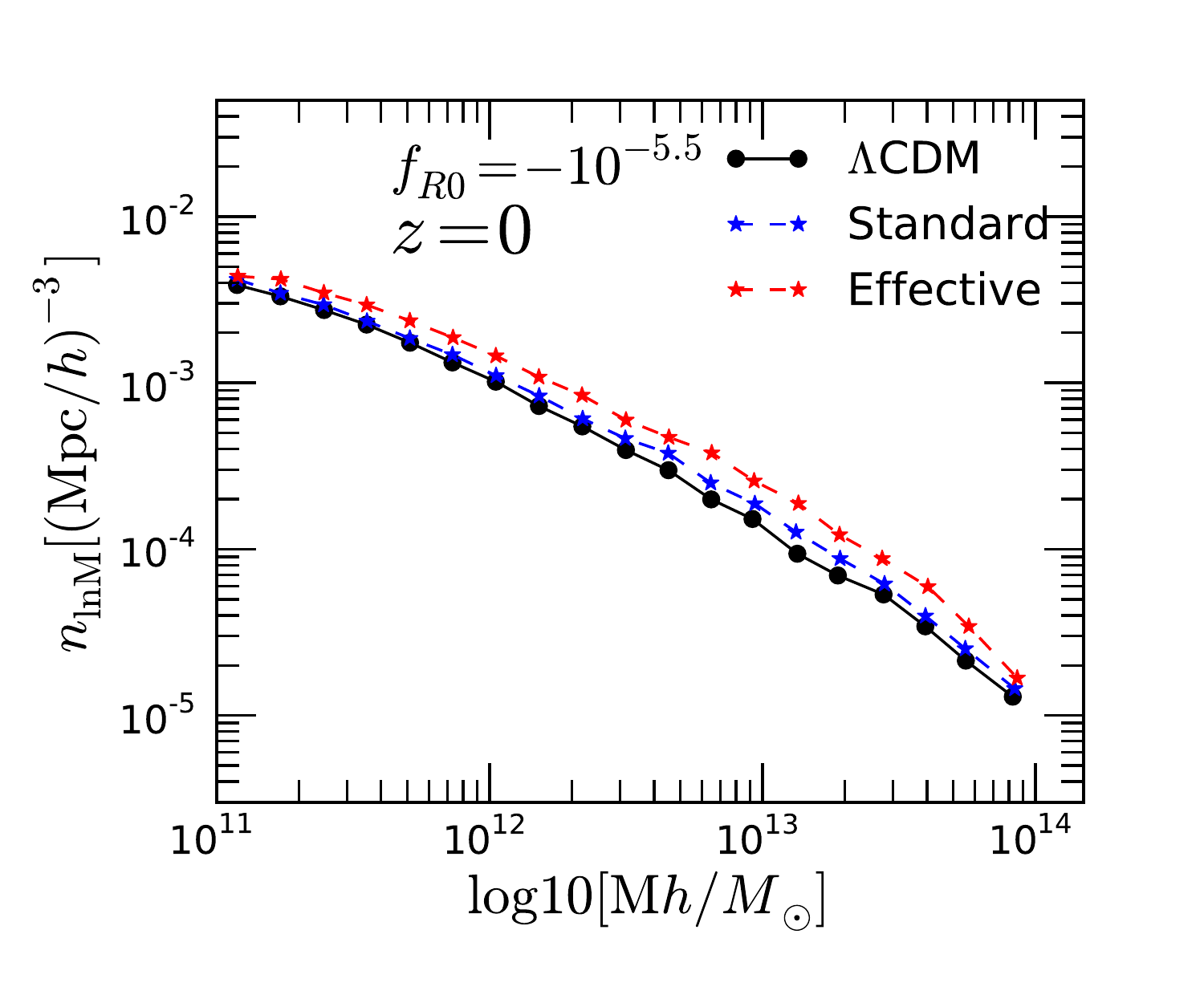}
\caption{Mass function of the effective and standard halos compared with that of the $\Lambda$CDM model. $n_{\ln M}\equiv\frac{dn}{d\ln M}$ is the comoving number
density of halos per logarithmic interval in mass $M$. The effective halos show more significant enhancement in abundance than that of the standard halos.}\label{massfunc}
\end{figure}

{\bf Summary.} In the {\it standard} catalog of $f(R)$ gravity, halo properties %are dependent
depend not only on their masses but also on their level of screening. %The modification to gravity increases the complexity of halo properties.
However, by introducing the {\it effective} dark matter halo catalog, we find that the relationships between the {\it effective} mass of a halo and its dynamical properties, such as the density profile (or equivalently potential profile), velocity dispersion, specific angular momentum and spin, closely resemble those in $\Lambda$CDM cosmology. Thus, the {\it effective} dark matter halo catalog provides a convenient way for analyzing the baryonic physics in $f(R)$ gravity such as the X-ray cluster and the Large-Scale Sunyaev-Zel'dovich effect~\cite{SZ,SZ_lcdm}. Since the effective halos show more significant enhancement in their abundance than that of the standard halos, the measurement of galaxy cluster abundance can provide more robust constraints on $f(R)$ gravity than using the standard halos (e.g. Ref.~\cite{cluster}.)  The {\it effective} dark matter halo catalog can also give us some basic insights into the galaxy formation in $f(R)$ gravity. On the scale of galaxies, the self-gravity of gas (which is the dominant baryonic
component of galaxies) can be neglected,
%Since on such scales, the self-gravity of gas can be neglected,
and so ``gastrophysics" such as cooling and the accretion of gas, in an {\it effective} halo in $f(R)$ gravity should follow the same relationship with mass as halos in a $\Lambda$CDM cosmology. It can therefore be expected that galaxy halo occupation distribution models, designed to work in a $\Lambda$CDM cosmology, can be straightforwardly applied to the {\it effective} halo catalog in $f(R)$ gravity. Furthermore, although we demonstrated this idea for a specific $f(R)$ gravity model, it can be generalised to other modified gravity or coupled dark energy models, and therefore is expected to have a much wider application.

%However, it should also be noted that
Of course, the baryonic physics inside galaxies can also be %subject to change under
changed by modified gravity, especially in regions where the self-gravity of baryons dominate over the dark matter. The processes such as the formation of stars and feedback might be sensitive to the modification of gravity, and as a result %onsequently,
galaxy properties such as color and luminosity may differ from the $\Lambda$CDM predictions. %in the standard gravity, this could actually open the door to a novel way of testing gravity. However, in order to draw quantitative conclusions, we need to investigate this baryonic physics and build self consistent semi-analytical galaxy formation models. We shall address this in future work. %Further, although we illustrate the idea of effective halo catalog within $f(R)$ gravity, this method can be generalized to other modified gravity theories as well.
Although further studies on these topics are needed, we have seen that the introduction of {\it effective} halo catalog can greatly simplify the analysis of physical processes of galaxy formation in modified gravity and therefore will be a useful initial step in this direction.

{\bf acknowledgments}
JHH acknowledges support of the Italian Space Agency (ASI), via contract agreement I/023/12/0. AJH and LG are supported by the European Research
Council through the Darklight ERC Advanced Research Grant (291521).

\end{document}